\documentclass{aa}
\usepackage{natbib}
\usepackage{txfonts}
\usepackage{hyperref}
\usepackage{graphicx}
\usepackage{subfigure}
\graphicspath{{figures/}}

\begin{document}
 
\title{Period spacing of gravity modes strongly affected by rotation}
\subtitle{Going beyond the traditional approximation}

\author{V. Prat \inst{1,2} \and S. Mathis\inst{1} \and F. Ligni\`eres\inst{3,4} \and J. Ballot\inst{3,4} \and P.-M. Culpin\inst{3,4}}
\institute{
Laboratoire AIM Paris-Saclay, CEA/DRF - CNRS - Universit\'e Paris Diderot, IRFU/SAp Centre de Saclay, F-91191 Gif-sur-Yvette, France
\and
Max-Planck Institut f\"ur Astrophysik, Karl-Schwarzschild-Str. 1, 85748, Garching bei M\"unchen, Germany
\and
Universit\'e de Toulouse; UPS-OMP; IRAP; Toulouse, France
\and
CNRS; IRAP; 14 avenue \'Edouard Belin; F-31400 Toulouse, France}

\date{}

\abstract
{As of today, asteroseismology mainly allows us to probe the internal rotation of stars when modes are only weakly affected by rotation using perturbative methods.
Such methods cannot be applied to rapidly rotating stars, which exhibit complex oscillation spectra.
In this context, the so-called traditional approximation, which neglects the terms associated with the latitudinal component of the rotation vector, describes modes that are strongly affected by rotation. This approximation is sometimes used for interpreting asteroseismic data,
however, its domain of validity is not established yet.}
{We aim at deriving analytical prescriptions for period spacings of low-frequency gravity modes strongly affected by rotation through the full Coriolis acceleration (i.e. without neglecting any component of the rotation vector), which can be used to probe stellar internal structure and rotation.}
{We approximated the asymptotic theory of gravito-inertial waves in uniformly rotating stars using ray theory described in a previous paper in the low-frequency regime, where waves are trapped near the equatorial plane.
We put the equations of ray dynamics  into a separable form and used the Einstein-Brillouin-Keller (EBK) quantisation method to compute modes frequencies from rays.}
{Two spectral patterns that depend on stratification and rotation are predicted within this new approximation: one for axisymmetric modes and one for non-axisymmetric modes.}
{The detection of the predicted patterns in observed oscillation spectra would give constraints on internal rotation and chemical stratification of rapidly rotating stars exhibiting gravity modes, such as $\gamma$ Doradus, SPB, or Be stars.
The obtained results have a mathematical form that is similar to that of the traditional approximation, but the new approximation takes the full Coriolis, which allows for propagation near the centre, and centrifugal accelerations into account.}


\maketitle

\section{Introduction}

Rotation is a key process of the evolution of stars.
Indeed, in radiative zones rotation simultaneously generates large-scale motions, such as meridional circulation, and small-scale turbulent motions, which are induced by hydrodynamical instabilities such as the shear instability.
Along with the magnetic field, rotation strongly affects the transport of chemical elements and angular momentum, and thus the structure and evolution of stars \citep{Zahn, Spruit, Maeder, Mathis13}.

Surface rotation can be estimated with various methods, including spectrometry \citep[e.g.][]{Ramirez}, spectro-polarimetry \citep[e.g.][]{Paletou}, spectro-interferometry \citep[e.g.][]{Hadjara}, and photometry \citep[e.g.][]{Garcia14}.
In contrast, internal rotation is much harder to constrain.
In this context, the development of helio- and asteroseismology \citep{Aerts10}, and the high-quality data provided by the space missions CoRoT \citep{Baglin} and \emph{Kepler} \citep{Borucki} give access to the physics of stellar interiors, including internal rotation.

For slow rotators, such as solar-type stars and red giants, rotation is usually considered a perturbation of a non-rotating system that generates splittings between modes of different azimuthal orders \citep{Saio}.
The differential rotation in the Sun has been constrained using rotational splittings of \emph{p} modes \citep{Thompson, Couvidat} down to $0.2R_\odot$ and of \emph{g}-mode candidates \citep{Garcia07}.
The contrast in rotation between the core and surface has also been measured for subgiant and red giant stars using rotational splittings of mixed modes \citep{Beck, Mosser, Deheuvels12, Deheuvels14, Deheuvels15}.
Recently, similar techniques have been applied to solar-type stars \citep{Benomar} and a few F-, A-, and B-type stars \citep{Aerts, Kurtz, Saio15, Triana}.

The picture is different for rapidly rotating intermediate-mass and massive pulsators, such as $\gamma$ Doradus, $\delta$ Scuti, SPB, $\beta$ Cephei, or Be stars.
Our understanding of the effects of rotation on oscillation modes is still limited.
Indeed, perturbative methods are not valid for high rotation rates \citep{Reese06, Ballot10, Ballot13}.
Another approximation, namely the traditional approximation \citep{Eckart}, can be used to simplify the effect of the Coriolis acceleration on low-frequency \emph{g} modes \citep{LeeSaio,Townsend03, Bouabid, Ouazzani}.
This approximation consists in neglecting the latitudinal component of the rotation vector in the Coriolis acceleration.
If one further neglects the centrifugal deformation, the problem of finding eigenmodes then becomes separable in the spherical coordinates.
Computations within this approximation are used by \citet{Moravveji} and \citet{VanReeth} to constrain internal rotation and mixing processes in rapidly rotating stars.
However, the relevance of this approximation for stellar oscillations is uncertain \citep[see for example][]{Friedlander,Gerkema, Mathis14, PratLB}.

Thus, methods are needed that are able to describe consistently the full effects of rotation on the oscillation spectra when the rotation is not a perturbation.
Indeed, except in some special cases \citep[see e.g.][for inertial waves in a spheroid]{Bryan}, the problem is fully bidimensional, that is not separable.
The first non-traditional calculations in stellar physics have been carried out by \citet{DintransRieutord}, and generalised to differentially rotating cases by \citet{Mirouh}.
A promising way of elaborating such methods is to build asymptotic theories of modes based on the short-wavelength approximation, which describes the propagation of waves using ray models similar to geometrical optics.
Such a theory has been first built for high-frequency acoustic modes \citep{LG08, LG09, Pasek11, Pasek12}.

Concerning gravito-inertial modes, a first step in this direction has been achieved by \citet{PratLB} (hereafter refered to as Paper I).
We have derived a general eikonal equation (i.e. a local dispersion relation) of gravito-inertial waves in a uniformly rotating, centrifugally deformed star and the equations governing ray dynamics.
The structure of the phase space has been investigated thanks to a ray-tracing code.
In particular, the phase space is nearly integrable at low frequencies as the dominant structures are invariant tori.

The purpose of the present paper is to derive an approximate version of the ray dynamics that is integrable to analytically compute the mode frequencies and propose simple prescriptions for period spacings that go beyond the traditional approximation.
A summary of the ray-based asymptotic theory of Paper I is presented in Sect.~\ref{sec:sum}.
The low-frequency approximation used to simplify the dynamics is derived in Sect.~\ref{sec:lfa}.
Period spacings based on the approximate dynamics are proposed in Sect.~\ref{sec:dia}.
Then, the new formalism is compared with the traditional approximation in Sect.~\ref{sec:comp}.
Finally, we discuss the range of validity of our results and conclude in Sect.~\ref{sec:dis}.

\section{Summary of the asymptotic theory}
\label{sec:sum}

\subsection{General case}
\label{sec:gen}

According to Paper I, the general eikonal equation (i.e. local dispersion relation) for non-axisymmetric gravito-inertial waves in a uniformly rotating star is written
\begin{equation}
    \label{eq:eik}
    \omega^2 = \frac{f^2k_z^2+N_0^2\left(k_\perp^2+k_\phi^2\right)+f^2\cos^2\Theta k_{\rm c}^2}{k^2+k_{\rm c}^2},
\end{equation}
where $\omega$ is the pulsation in the rotating frame; $f=2\Omega$, where $\Omega$ is the rotation rate; $N_0$ is the Brunt-V\"ais\"al\"a frequency, which is the upper bound for frequencies of pure gravity waves; $k_z$, $k_\perp$, $k_\phi$, and $k$ are the component parallel to the rotation axis, the component perpendicular to the effective gravity (which takes the centrifugal acceleration into account) in the meridional plane, the azimuthal component, and the norm of the wave vector $\vec k$, respectively; $\Theta$ is the angle between the rotation axis and the direction opposite to the effective gravity; $k_{\rm c}$ is a term that becomes dominant near the surface and is responsible for the back refraction of waves.

Equation~\eqref{eq:eik} was derived in Paper~I assuming a polytropic background stellar model, but in fact, only the surface term is affected by this assumption.
For more realistic background models, one might want to find a more general form for this term (see Appendix~A.3 of Paper~I), but the polytropic version is sufficient to refract waves back into the star, so we choose to keep it here, even without assuming a polytropic model for $N_0$.

In the axisymmetric case ($k_\phi=0$) and far from the surface (where $k_{\rm c}$ can be neglected), the eikonal equation~\eqref{eq:eik} can be seen as a quadratic equation in any meridional component of the wave vector, say $k_z$.
The propagation of waves requires the existence of real solutions, which means that the discriminant of the equation has to be positive.
This reduces to the condition $\Gamma\geq0$, where
\begin{equation}
    \label{eq:gam}
    \Gamma=-\omega^4+\omega^2(f^2+N_0^2)-N_0^2f^2\cos^2\Theta.
\end{equation}
Near the centre of the star, $N_0$ vanishes, and the condition of propagation simplifies into $\omega\leq f$.
In the corresponding regime, called sub-inertial, waves can propagate near the centre of the star.
Besides, in regions where $N_0\gg f$, the same condition yields $\omega\geq f|\cos\Theta|$, which means that sub-inertial waves are trapped in the equatorial region (see Fig.~2b of Paper~I).

\begin{figure}
    \resizebox{\hsize}{!}{\includegraphics{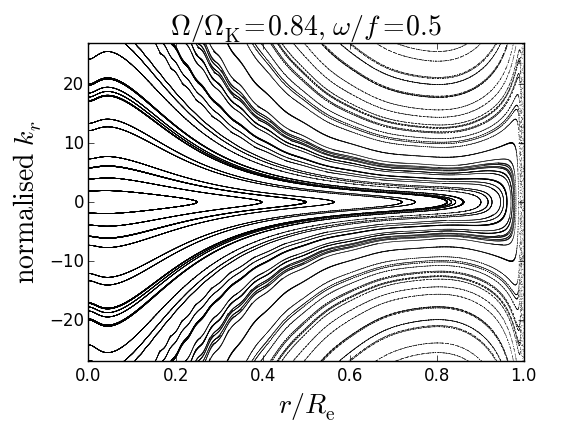}}
    \caption{PSS of a polytropic model of star (with a polytropic index $\mu=3$) rotating at 84\% of its critical velocity $\Omega_{\rm K}=\sqrt{GM/R_{\rm e}^3}$, where $M$ is the mass of the star and $R_{\rm e}$ its equatorial radius, where $\omega/f=0.5$.}
    \label{fig:pss}
\end{figure}

The equations governing the ray dynamics are derived from the eikonal equation \eqref{eq:eik} using the relations
\begin{align}
    \frac{{\rm d} x_i}{{\rm d}t} &= \frac{\partial\omega}{\partial k_i}, \\
    \frac{{\rm d} k_i}{{\rm d}t} &= -\frac{\partial\omega}{\partial x_i},
\end{align}
where $k_i=\partial\Phi/\partial x_i$ is the covariant component of the wave vector in the natural basis $\vec e_i=\partial\vec x/\partial x_i$ associated with the variable $x_i$ ($\Phi$ is the phase of the wave).
These relations ensure that the pulsation $\omega$ is invariant along the ray trajectories.
In Paper I, we only investigated the axisymmetric case numerically.
An overview of the nature of the dynamics can be obtained by looking at Poincaré surfaces of section (PSS).
A PSS is the intersection of all ray trajectories with a given surface, here the equatorial plane.
An example of PSS in the low-frequency regime is shown in Fig.~\ref{fig:pss}.
In particular, one can see that in this regime, the phase space is mostly made of invariant tori that intersect the PSS on unidimensional curves.
The evolution of the phase space with increasing rotation shows that these tori result from a smooth deformation of the non-rotating tori and that they persist even for large centrifugal deformations.
This suggests the existence of a nearby integrable system.

\subsection{Traditional approximation}
\label{sec:tradi}

One may think that such a system can be obtained using the so-called traditional approximation.
This approximation consists in neglecting the centrifugal deformation and assuming that in the Coriolis acceleration, only the radial part of the rotation vector $\vec\Omega$ matters (i.e. $\vec\Omega\simeq\Omega\cos\theta\vec e_r$, where $\theta$ is the colatitude and $\vec e_r$ the radial unit vector).
This implies the separability of eigenmodes using the so-called Hough functions in latitude, which reduce to Legendre polynomials in the non-rotating case \citep[e.g.][]{LeeSaio}.

We investigated the impact of the traditional approximation on the ray dynamics in Paper~I.
In particular, the eikonal equation written in spherical coordinates becomes
\begin{equation}
    \label{eq:eik_tra}
    \omega^2 = \frac{f^2\cos^2\theta\left(k_r^2+k_{\rm c}^2\right)+N_0^2\left(k_\theta^2+k_\phi^2\right)}{k^2+k_{\rm c}^2},
\end{equation}
where $k_r$ and $k_\theta$ are the components of the wave vector in the spherical coordinates $(r,\theta)$.
The corresponding condition of propagation in the traditional approximation is
\begin{equation}
   (\omega^2-f^2\cos^2\theta)(N_0^2-\omega^2) \geq 0. 
\end{equation}
This implies that near the centre, sub-inertial waves in the traditional approximation can only propagate near the rotation axis.
As we have shown in Sect.~4.2 of Paper~I, it is possible to separate the ray dynamics into radial and latitudinal parts and to find a second invariant, which allows one to describe the structure of the phase space analytically.

\section{Low-frequency approximation}
\label{sec:lfa}

In this section we search for such an integrable system that would allow us to find an analytical prescription for the mode frequencies while taking the full Coriolis acceleration (without neglecting any component of the rotation vector) into account.
The first step is to use the assumption that the frequency is much lower than the Coriolis frequency to simplify the eikonal equation, as explained in Sect.~\ref{sec:eik}.
Then, one can deduce the existence of a second invariant from the corresponding ray dynamics, as shown in Sect.~\ref{sec:ray}.
Finally, the Einstein-Brillouin-Keller (EBK) quantisation method is used to describe the frequency spectrum that can be constructed from rays (Sect.~\ref{sec:mod}).

\subsection{Approximation of the eikonal equation}
\label{sec:eik}

One of the problems raised by Eq.~\eqref{eq:eik} is that the numerator involves $k_z$ and $k_\perp$, the components of the wave vector along $\vec e_z$ and $\vec e_\perp$, which do not form an orthonormal frame.
When writing the eikonal equation in the cylindrical or the spherical frame, mixed terms with products of different components of $\vec k$ appear.
For example, Eq.~\eqref{eq:eik} becomes in the cylindrical frame
\begin{equation}
    \label{eq:eik_cyl}
    \omega^2 = \frac{
    \begin{aligned}
        &\left[\left(f^2+N_0^2\sin^2\Theta\right)k_z^2-2N_0^2\sin\Theta\cos\Theta k_sk_z\right.   \\
        &\left.\quad+N_0^2\cos^2\Theta k_s^2+N_0^2k_\phi^2+f^2\cos^2\Theta k_{\rm c}^2\right]
    \end{aligned}
    }{k^2+k_{\rm c}^2},
\end{equation}
where $k_s$ is the component of the wave vector in the direction orthogonal to the rotation axis.

It is possible to get rid of the mixed term by introducing a new orthonormal frame given by
\begin{align}
    k_z &= k_\beta\cos\alpha - k_\gamma\sin\alpha,  \label{eq:pri_kz}   \\
    k_s &= k_\beta\sin\alpha + k_\gamma\cos\alpha,  \label{eq:pri_ks}
\end{align}
where the angle $\alpha$ is chosen in such a way that the mixed term vanishes.
This condition is written
\begin{equation}
    \label{eq:def_pri}
    N_0^2\sin[2(\alpha-\Theta)] = f^2\sin 2\alpha.
\end{equation}
The eikonal equation~\eqref{eq:eik_cyl} then becomes
\begin{equation}
    \label{eq:eik_pri}
    \omega^2 = \frac{
    \begin{aligned}
        &\left\{\left[f^2\cos^2\alpha+N_0^2\sin^2(\alpha-\Theta)\right]k_\beta^2 +f^2\cos^2\Theta k_{\rm c}^2\right.
   \\
        &\left.\quad+\left[f^2\sin^2\alpha+N_0^2\cos^2(\alpha-\Theta)\right]k_\gamma^2 +N_0^2k_\phi^2\right\}
    \end{aligned}
    }{k^2+k_{\rm c}^2}.
\end{equation}

In the low-frequency regime defined by $\omega\ll f$, the first term of $\Gamma$ in Eq.~\eqref{eq:gam} can be neglected and the condition of propagation for axisymmetric waves becomes
\begin{equation}
    \label{eq:ineq}
    \frac{N_0^2}{N_0^2+f^2}\cos^2\Theta \leq \frac{\omega^2}{f^2} \ll 1,
\end{equation}
which implies that those waves are trapped near the equatorial plane.
Similarly to what was explained in Sect.~\ref{sec:gen}, for non-axisymmetric waves and when surface effects are neglected, Eq.~\eqref{eq:eik_cyl} can be rewritten as a quadratic equation in $k_z$ as follows:
\begin{equation}
    \begin{aligned}
        \left(f^2+N_0^2\sin^2\Theta-\omega^2\right)k_z^2-2N_0^2\sin\Theta\cos\Theta k_sk_z  \\
        \quad+\left(N_0^2\cos^2\Theta-\omega^2\right)k_s^2+\left(N_0^2-\omega^2\right)k_\phi^2=0.
    \end{aligned}
\end{equation}
The reduced discrimant of this equation,
\begin{equation}
    \label{eq:disc}
    \Delta'= \left(k_s^2+k_\phi^2\right)\Gamma - k_\phi^2N_0^2\sin^2\Theta\left(N_0^2+f^2-\omega^2\right),
\end{equation}
has to be positive for $k_z$ to be real and for waves to propagate.
For sub-inertial waves, $N_0^2+f^2-\omega^2$ is always positive.
The propagation condition $\Delta'\geq0$ thus implies that $\Gamma\geq0$.
This means that Eq.~\eqref{eq:ineq} is also verified for low-frequency, non-axisymmetric waves and that those are also trapped near the equatorial plane.

We now use this trapping, through Eq.~\eqref{eq:ineq}, to simplify the canonical eikonal equation~\eqref{eq:eik_pri}.
Choosing the solution of Eq.~\eqref{eq:def_pri} for which $k_\beta=k_s$ at the equator, we finally obtain (see Appendix~\ref{app:lfa})
\begin{equation}
    \label{eq:eik_gam}
    \omega^2 = \frac{f^2\cos^2\delta\left(k_\beta^2 + k_{\rm c}^2\right) + \left(N_0^2+f^2\right)k_\gamma^2+N_0^2k_\phi^2}{k^2+k_{\rm c}^2},
\end{equation}
where $\delta$ is defined by
\begin{equation}
    \label{eq:del}
    \cos^2\delta = \frac{N_0^2}{N_0^2+f^2}\cos^2\Theta.
\end{equation}

\subsection{Separable ray dynamics}
\label{sec:ray}

\begin{figure}
    \resizebox{\hsize}{!}{\includegraphics{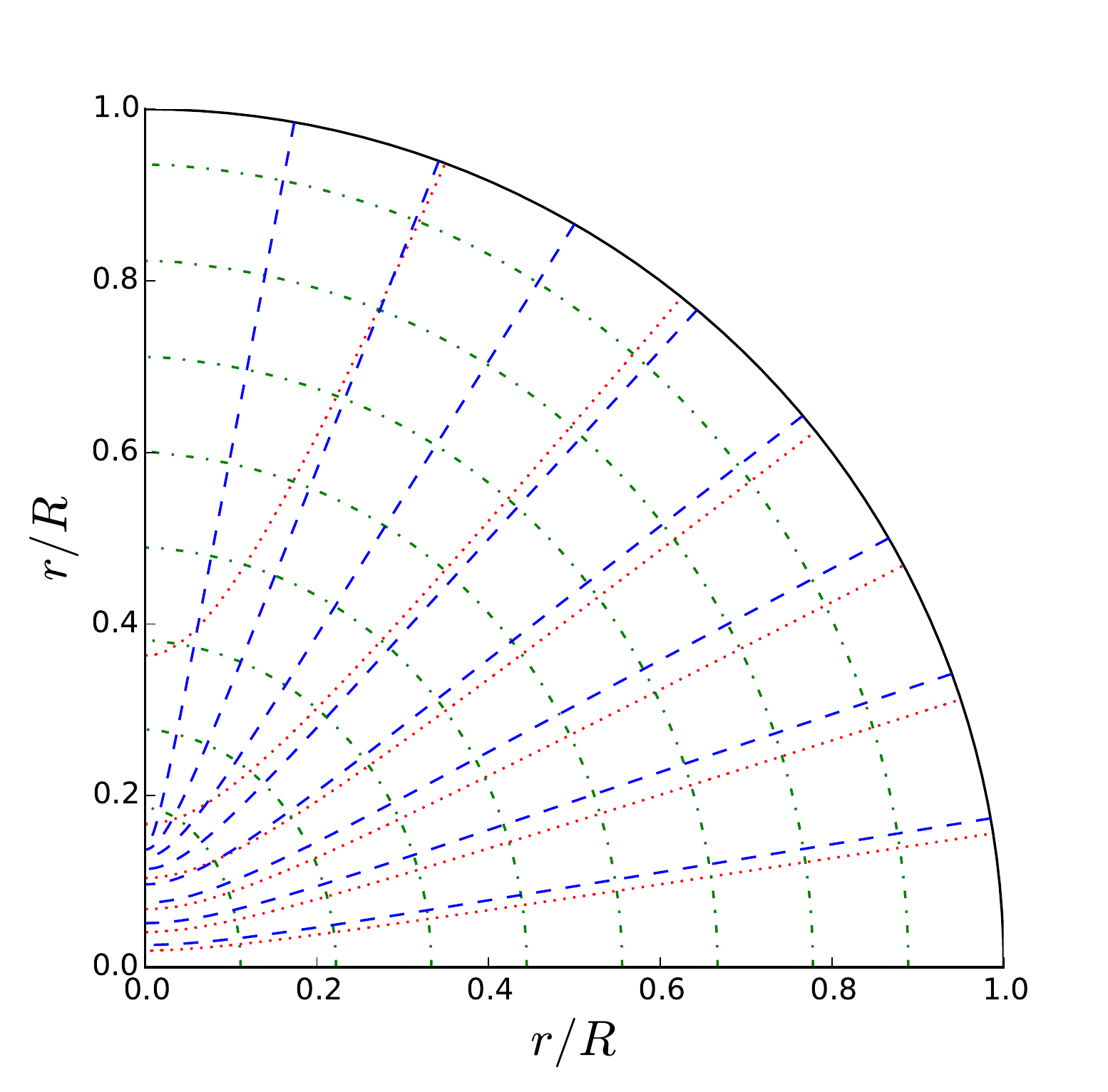}}
    \caption{New coordinate system in the meridional plane of a spherical star of radius $R$ with $N_0(r)=4f\exp(-2r/R)$. Green dash-dotted lines, red dotted lines, and blue dashed lines are iso-contours of $\beta$, $\delta$, and $\gamma$, as defined in Eqs.~\eqref{eq:dbet}, \eqref{eq:del}, and \eqref{eq:dgam}, respectively.}
    \label{fig:map}
\end{figure}

To benefit from the simplified version of the eikonal equation and write the corresponding equations of the ray dynamics, one first has to define the coordinate system ($\beta$, $\gamma$) associated with the components $k_\beta$ and $k_\gamma$ of the wave vector. 
Using Eqs.~\eqref{eq:pri_kz} and \eqref{eq:pri_ks} and the fact that in a natural coordinate system, one has $k_i=\partial\Phi/\partial x_i$, where $\Phi$ is the phase of the wave, we write
\begin{align}
    {\rm d}\beta &= {\rm d}s \sin\alpha + {\rm d}z\cos\alpha,   \label{eq:dbet} \\
    {\rm d}\gamma &= {\rm d}s \cos\alpha - {\rm d}z\sin\alpha.  \label{eq:dgam}
\end{align}
These new coordinates are illustrated in Fig.~\ref{fig:map}.

When close to the equator, we observe that surfaces of constant $\gamma$ are close to those of constant $\delta$.
Besides, Eq.~\eqref{eq:eik_gam} is written in terms of $\delta$.
Thus, we choose to replace $k_\gamma$ by its expression as a function of $k_\delta$, which is the natural component of the wave vector associated with $\delta$.
With some approximations explained in Appendix~\ref{sec:gamdel}, one can write
\begin{equation}
    \label{eq:kdel}
    k_\gamma\simeq\frac{k_\delta}{\zeta},
\end{equation}
where $\zeta$ is defined by
\begin{equation}
    \label{eq:zet}
    \zeta = \frac{r\sqrt{N_0^2+f^2}}{N_0}.
\end{equation}
Similarly, $k_\phi$ can be expressed as a function of the natural component of the wave vector associated with $\phi$, that is
\begin{equation}
    k_\phi = \frac{m}{s},
\end{equation}
where $m$ is the number of nodes in the azimuthal direction.
Since $\sin\theta\simeq 1$ near the equator, one can further write
\begin{equation}
    k_\phi \simeq \frac{m}{r}.
\end{equation}
The final eikonal equation expressed in the coordinate system ($\beta$, $\delta$) is then
\begin{equation}
    \label{eq:eik_del}
    \omega^2 = \frac{f^2\cos^2\delta\left(k_\beta^2 + k_{\rm c}^2\right) + \left(N_0^2+f^2\right)\frac{k_\delta^2+m^2}{\zeta^2}}{k^2+k_{\rm c}^2}.
\end{equation}

To go further, we make the coarse approximation that $N_0$, $k_{\rm c}$, and $\zeta$ depend only on $\beta$.
It is rigourously true only in the non-rotating case, but because waves are trapped near the equatorial plane, the variations of those quantities with $\delta$ may be neglected.
Equation~\eqref{eq:eik_del} then yields
\begin{align}
    \frac{{\rm d}\beta}{{\rm d}t}       &= \frac{\left(f^2\cos^2\delta - \omega^2\right)k_\beta}{\omega(k^2+k_{\rm c}^2)}, \label{eq:dynbet}\\
    \frac{{\rm d}k_\beta}{{\rm d}t}     &= \frac{
        \openup-2\jot
        \begin{aligned}
            &\biggl\{\left[(N_0^2)'\zeta^2-(\zeta^2)'\left(N_0^2+f^2-\omega^2\right)\right]\tfrac{k_\delta^2+m^2}{\zeta^4}\biggr.    \\
            &\biggl.\quad+(k_{\rm c}^2)'\left(f^2\cos^2\delta - \omega^2\right)\biggr\}
        \end{aligned}
    }{\omega(k^2+k_{\rm c}^2)}, \label{eq:dynkbet}\\
    \frac{{\rm d}\delta}{{\rm d}t}      &= \frac{\left(N_0^2+f^2-\omega^2\right)k_\delta}{\omega\zeta^2(k^2+k_{\rm c}^2)}, \label{eq:dyndel}\\
    \frac{{\rm d}k_\delta}{{\rm d}t}    &= \frac{f^2\sin\delta\cos\delta\left(k_\beta^2+k_{\rm c}^2\right)}{\omega(k^2+k_{\rm c}^2)},   \label{eq:dynkdel}
\end{align}
where $'$ denotes the derivative with respect to $\beta$.

We deduce from these equations that the quantity
\begin{equation}
    \label{eq:chi}
    \chi = \frac{N_0^2+f^2\sin^2\delta}{\zeta^2(k^2+k_{\rm c}^2)}
\end{equation}
is an invariant of the dynamics (see Appendix~\ref{sec:invar}).
The existence of this second invariant (in addition to the pulsation $\omega$) proves that the considered system is integrable.

The new invariant allows us to write the components of the wave vector as functions of the two invariants and of the spatial coordinates,
\begin{align}
    k_\beta^2 + k_{\rm c}^2 &= \frac{N_0^2+f^2-\omega^2}{\zeta^2\chi},  \label{eq:kbe}  \\
    k_\delta^2              &= \frac{\omega^2-f^2\cos^2\delta}{\chi}-m^2.   \label{eq:kde}
\end{align}
The system is thus separable in $\beta$ and $\delta$ in the sense that spatially, $k_\beta$ (resp. $k_\delta$) depends only on $\beta$ (resp. $\delta$).

\begin{figure}
    \resizebox{\hsize}{!}{\includegraphics{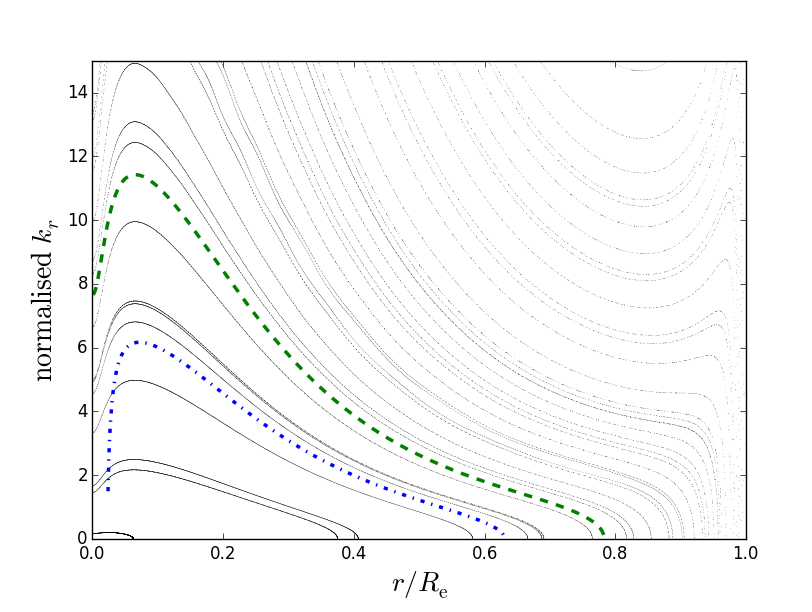}}
    \caption{PSS of a polytropic model of star (with a polytropic index $\mu=3$) rotating at 38\% of its critical velocity with $\omega/f=0.8$. The green dashed and blue dash-dotted lines are imprints of trajectories in the low-frequency approximation and in the traditional approximation, respectively.}
    \label{fig:lf_lr}
\end{figure}

\begin{figure}
    \resizebox{\hsize}{!}{\includegraphics{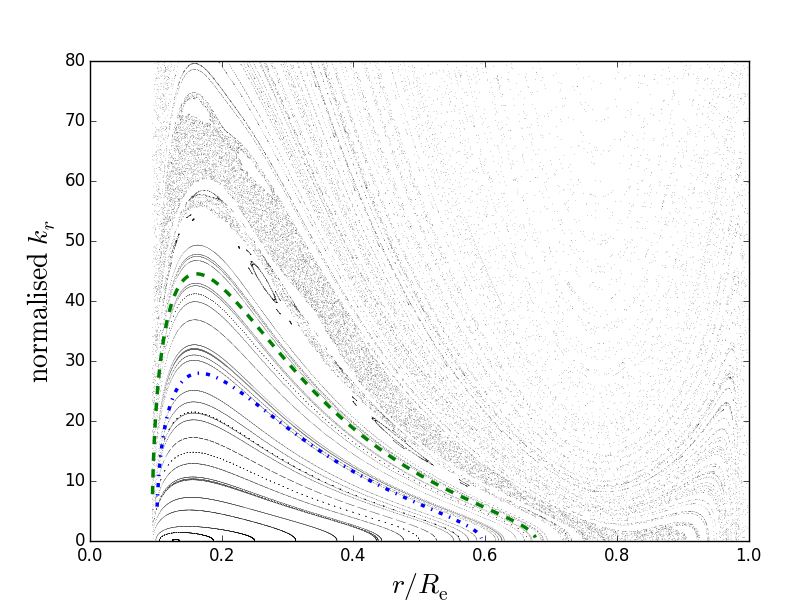}}
    \caption{Same as Fig.~\ref{fig:lf_lr}, but with $\omega/f=3.1$.}
    \label{fig:hf_lr}
\end{figure}

\begin{figure}
    \resizebox{\hsize}{!}{\includegraphics{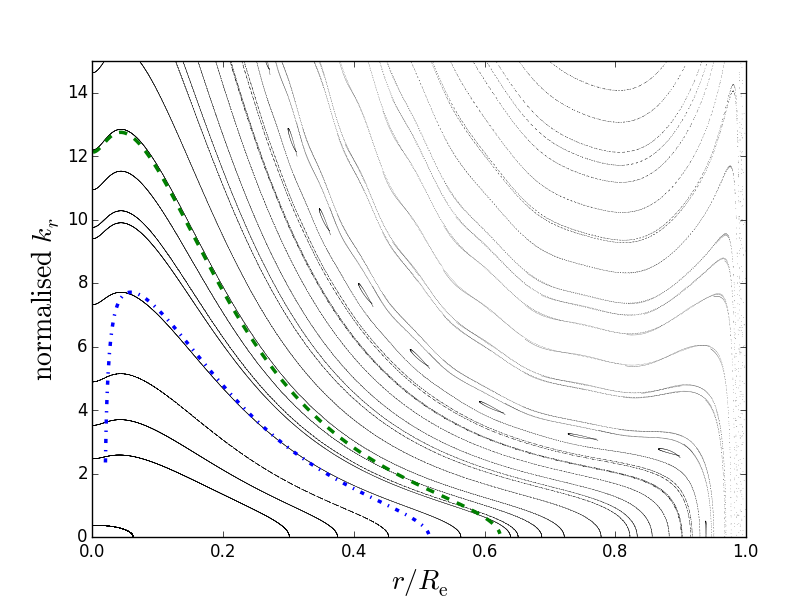}}
    \caption{Same as Fig.~\ref{fig:lf_lr}, but with a rotation of 84\% of the critical velocity and $\omega/f=0.5$.}
    \label{fig:lf_hr}
\end{figure}

Since $k_\beta=k_r$ on the equatorial plane, Eq.~\eqref{eq:kbe} can be used to compute analytically the imprints of the rays considered here on PSS, such as those presented in Paper~I.
As shown in Fig.~\ref{fig:lf_lr}, the low-frequency approximation (green dashed line) works well in the sub-inertial regime, including near the centre.
Surprisingly, it also works rather well in the super-inertial regime far from large island chains and chaotic regions, as illustrated in Fig.~\ref{fig:hf_lr}.
However, when rotation increases, the low-frequency approximation becomes less accurate (see Fig.~.\ref{fig:lf_hr}).
This is likely because the dependence of structure quantities, such as $N_0$ and $k_{\rm c}$, on $\delta$ can no longer be neglected when the star becomes fully two-dimensional \citep[see for example models described by][]{EspinosaRieutord}.

Even if looking at PSS is insightful, it is not sufficient to test the validity of the low-frequency approximation.
To do so, one may check how the new invariant is conserved in the full ray dynamics.
This gives the following results: the invariant is typically conserved up to 10\% in most of the star, except in a very thin region near the surface and in a small region between the core, where $N_0$ is much smaller than $f$, and the envelope, where it is much larger.

\subsection{From rays to modes}
\label{sec:mod}

Modes are formed by positive interference of propagating waves, which are described here  by rays.
To interfere positively with itself, the phase of a wave has to vary by a multiple of $2\pi$ after going back to the same point.
This is formalised by the quantisation condition
\begin{equation}
    \label{eq:quant}
    \int_\mathcal{C}\vec k\cdot{\rm d}\vec x = 2\pi\left(p+\frac{\varepsilon}{4}\right),
\end{equation}
where $\mathcal C$ is a closed curve formed by a ray trajectory, $p$ and $\varepsilon$ are integers, and $\varepsilon$ is the Maslov index accounting for phase shifts induced by boundaries \citep[see for example][and references therein]{LG09}.

\begin{figure}
    \resizebox{\hsize}{!}{\includegraphics{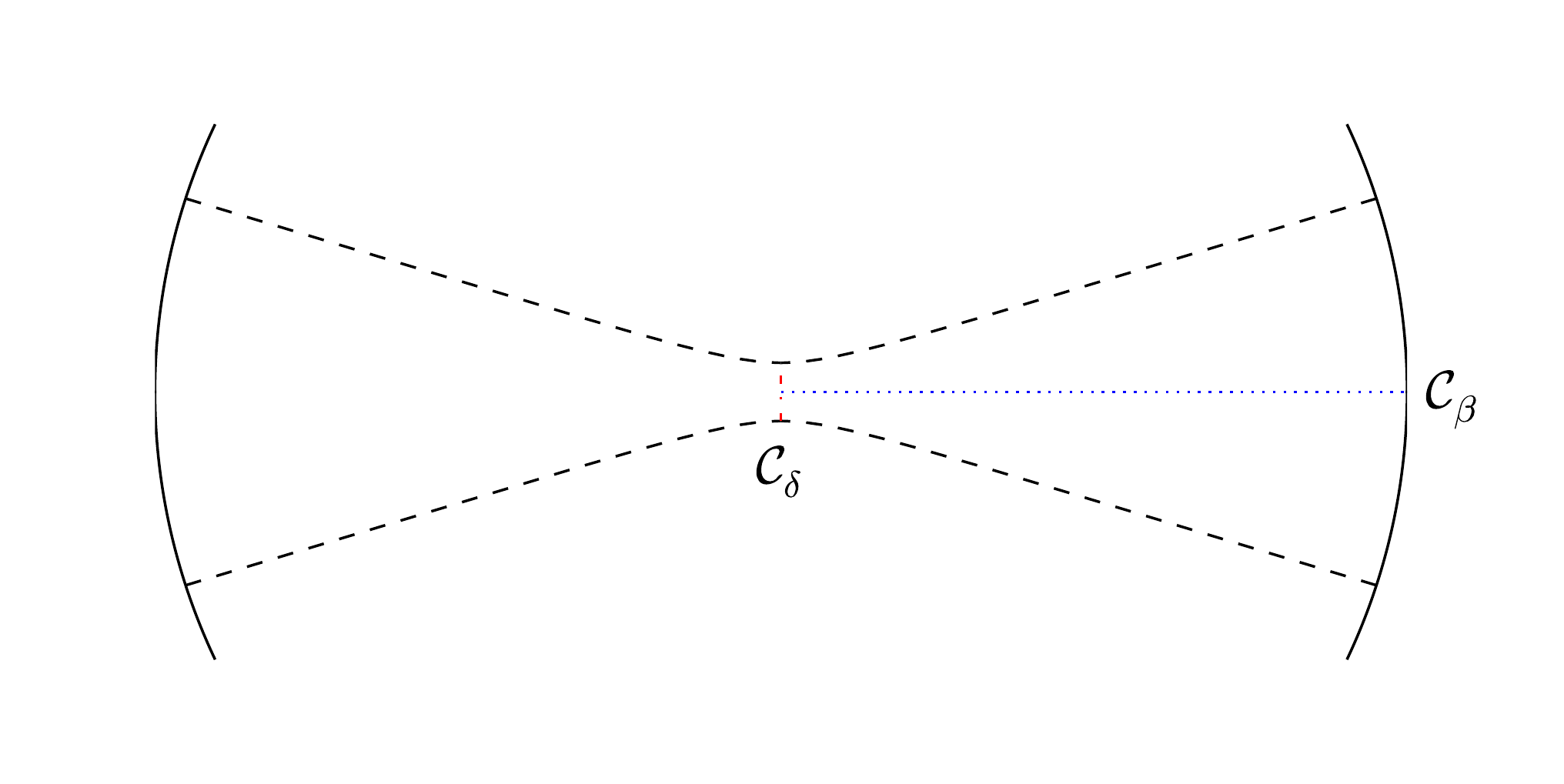}}
    \caption{Contours used for the quantisation. The black dashed lines correspond to an example of limits of the domain of propagation. The blue dotted line represents the contour $\mathcal{C}_\beta$ and the red dash-dotted line represents $\mathcal{C}_\delta$.}
    \label{fig:qua}
\end{figure}

For $N$-dimensional integrable systems, $N$ conditions can be obtained using $N$ independent closed curves, which do not necessarily correspond to rays.
Typically, one can choose curves defined by varying one coordinate and fixing the others.
In our case, it is particularly convenient to vary $\beta$ in the equatorial plane and $\delta$ along the rotation axis, as illustrated in Fig.~\ref{fig:qua}.
Equation~\eqref{eq:quant} then yields
\begin{align}
    \int_{\mathcal{C}_\beta}k_\beta{\rm d}\beta     &= 2\pi\left(\tilde n+\frac14\right),   \label{eq:qbet}\\
    \int_{\mathcal{C}_\delta}k_\delta{\rm d}\delta  &= 2\pi\left(\tilde\ell+\frac12\right), \label{eq:qdel}
\end{align}
where $\tilde n$ and $\tilde\ell$ are non-negative integers that are related to the radial order and to the degree of the mode, respectively.
By analogy with the non-rotating case, we have $\tilde n=n-1$ and $\tilde\ell=\ell_\mu-1$, where $n$ is the radial order and $\ell_\mu$ is the number of nodes in the latitudinal direction.

The contours $\mathcal{C}_\beta$ and $\mathcal{C}_\delta$ are closed curves, which means that the integrals in Eqs.~\eqref{eq:qbet} and \eqref{eq:qdel} can be split into two identical parts.
The symmetry of $\mathcal{C}_\delta$ with respect to the equatorial plane implies that the integral can be further split into four identical parts.
Using Eqs.~\eqref{eq:kbe} and~\eqref{eq:kde}, one obtains
\begin{align}
    \int_0^{r_{\rm s}}\sqrt{\frac{N_0^2+f^2-\omega^2}{\zeta^2\chi}-k_{\rm c}^2}{\rm d}r = \pi\left(\tilde n+\frac14\right),  \label{eq:qua_bet}\\
    \int_{\delta_{\rm c}}^{\frac{\pi}{2}}\sqrt{\frac{\omega^2-f^2\cos^2\delta}{\chi}-m^2}{\rm d}\delta = \frac{\pi}{2}\left(\tilde\ell+\frac12\right),    \label{eq:qua_del}
\end{align}
where $r_{\rm s}$ is the radius where $k_\beta$ vanishes and $\delta_{\rm c}$ is the limit value of $\delta$ given by
\begin{equation}
    \delta_{\rm c}=\arccos\frac{\sqrt{\omega^2-m^2\chi}}{f}.
\end{equation}
For $\delta_c$ to exist, $\omega^2-m^2\chi$ has to be positive.
In others words, one has the condition
\begin{equation}
    |m|\leq\frac{\omega}{\sqrt\chi}.
\end{equation}
The quantity $\lambda=\omega/\sqrt\chi$, which is an invariant, is the equivalent in the rotating case of $L=\ell+1/2$, where $\ell$ is the total number of nodes on the sphere.

In the general case, for given $\tilde n$, $\tilde\ell$, and $m$, Eqs.~\eqref{eq:qua_bet} and~\eqref{eq:qua_del} form a non-linear system of equations for $\omega$ and $\chi$.
It is possible, however, to simplify it.
First, in the low-frequency regime, $\omega$ can be neglected in Eq.~\eqref{eq:qua_bet}, and $r_{\rm s}$ can be considered independent of $\omega$.
Besides, if one considers that $k_{\rm c}$ has a negligible impact on the value of the integral, Eq.~\eqref{eq:qua_bet} yields
\begin{equation}
    \label{eq:sys_bet}
    \sqrt{\chi}=\frac{\int_0^{r_{\rm s}}\frac{N_0}{r}{\rm d}r}{\pi\left(\tilde n+\frac14\right)}.
\end{equation}
Using a Taylor expansion in $\sqrt{\omega^2-m^2\chi}/f=\cos\delta_{\rm c}$, Eq.~\eqref{eq:qua_del} can be rewritten
\begin{equation}
    \label{eq:sys_del}
    \omega^2 = m^2\chi + (2\tilde\ell+1)f\sqrt\chi.
\end{equation}
Expressed as a function of $\lambda$, one obtains
\begin{equation}
    \lambda^2 = (2\tilde\ell+1)\lambda\nu + m^2,
\end{equation}
where $\nu=f/\omega$ is the so-called spin factor.
This equation is very similar to what \citet{Townsend03} derived within the traditional approximation (see his Eqs.~[29] and [31]).
However, we miss one term, $m\nu$, because the term that accounts for Rossby and Kelvin waves has not been retained when deriving the eikonal equation in Paper~I.
This means that the present study cannot describe the corresponding modes.

Combining Eqs.~\eqref{eq:sys_bet} and~\eqref{eq:sys_del} yields
\begin{equation}
    \label{eq:ome}
    \omega^2 = \frac{(2\tilde\ell+1)f\int_0^{r_{\rm s}}\frac{N_0}{r}{\rm d}r}{\pi\left(\tilde n+\frac14\right)} + m^2\frac{\left(\int_0^{r_{\rm s}}\frac{N_0}{r}{\rm d}r\right)^2}{\pi^2\left(\tilde n+\frac14\right)^2}.
\end{equation}
The low-frequency assumption is verified when the two terms on the right-hand side   of Eq.~\eqref{eq:ome}
are much smaller than $f^2$.
This implies that both $2\tilde\ell+1$ and $m$ are much smaller than the quantity $\pi f(\tilde n+1/4)/\left(\int_0^{r_{\rm s}}N_0{\rm d}r/r\right)$, i.e. the radial order is much larger than the non-radial orders.

\section{Period spacing}
\label{sec:dia}

For non-axisymmetric modes, the observed frequency differs from the eigenfrequency computed in the rotating frame.
It is given by the relation
\begin{equation}
    \omega_{\rm obs} = \omega - m\Omega.
\end{equation}
Moreover, gravity modes are usually considered in terms of the observed period
\begin{equation}
    \Pi = \frac{2\pi}{\omega_{\rm obs}}.
\end{equation}
In general, the period of the mode characterised by the three numbers $\tilde n$, $\tilde\ell$, and $m$ is thus
\begin{equation}
    \label{eq:per}
    \Pi_{\tilde n\tilde\ell m} = \frac{2\pi}{\displaystyle{\sqrt{\frac{(2\tilde\ell+1)f\int_0^{r_{\rm s}}\frac{N_0}{r}{\rm d}r}{\pi\left(\tilde n+\frac14\right)} + m^2\frac{\left(\int_0^{r_{\rm s}}\frac{N_0}{r}{\rm d}r\right)^2}{\pi^2\left(\tilde n+\frac14\right)^2}}}-m\Omega}.
\end{equation}

\subsection{Axisymmetric modes}

We first consider the axisymmetric case.
Equation~\eqref{eq:per} simplifies into
\begin{equation}
    \Pi_{\tilde n\tilde\ell} = \sqrt{\frac{\pi^3\left(\tilde n+\frac14\right)}{\left(\tilde\ell+\frac12\right)\Omega\int_0^{r_{\rm s}}\frac{N_0}{r}{\rm d}r}}.
\end{equation}
The corresponding spectrum has regularities, but these are significantly different from the non-rotating case.
In particular, $\Pi_{\tilde n\tilde\ell}$ is proportional to $\sqrt{\tilde n+1/4}$, which means that the period spacing $\Delta\Pi = \Pi_{\tilde n+1\,\tilde\ell} - \Pi_{\tilde n\tilde\ell}$ is not constant.
For large values of $\tilde n$, one can instead write
\begin{equation}
    \label{eq:dpa}
    \Delta\Pi \simeq\frac{\pi^{3/2}}{\sqrt{2\left(\tilde n+\frac34\right)(2\tilde\ell+1)\Omega\int_0^{r_{\rm s}}\frac{N_0}{r}{\rm d}r}}.
\end{equation}
At a given $\tilde\ell$, the observation of period spacings, which scale as $1/\sqrt{\tilde n+3/4}$, would allow one to determine the value of the product $\Omega\int_0^{r_{\rm s}}N_0{\rm d}r/r$. 

\subsection{Non-axisymmetric modes}

The general expression given in Eq.~\eqref{eq:per} is not usable as such for giving simple prescriptions to interpret observations.
However, in the low-frequency regime, for non-zero $m$, $\omega$ is much smaller than $m\Omega$.
Using a Taylor expansion of Eq.~\eqref{eq:per} in $\omega/\Omega$, it yields
\begin{equation}
    \Pi_{\tilde n\tilde\ell m} \simeq -\frac{2\pi}{m\Omega}\left[1+\frac{\displaystyle{\sqrt{\frac{(2\tilde\ell+1)f\int_0^{r_{\rm s}}\frac{N_0}{r}{\rm d}r}{\pi\left(\tilde n+\frac14\right)} + m^2\frac{\left(\int_0^{r_{\rm s}}\frac{N_0}{r}{\rm d}r\right)^2}{\pi^2\left(\tilde n+\frac14\right)^2}}}}{m\Omega}\right].
\end{equation}
When computing the period spacing, the dominant term cancels out, and one obtains
\begin{equation}
    \label{eq:sep_m}
    \Delta\Pi \simeq \frac{2\int_0^{r_{\rm s}}\frac{N_0}{r}{\rm d}r}{m\Omega\left(\tilde n+\frac34\right)^2}\frac{1+\frac{\sigma}{2}}{\sqrt{1+\sigma}},
\end{equation}
where
\begin{equation}
    \sigma = \frac{(2\tilde\ell+1)f\pi\left(\tilde n+\frac34\right)}{m^2\int_0^{r_{\rm s}}\frac{N_0}{r}{\rm d}r} \gg \frac{2\tilde\ell+1}{m}.
\end{equation}

Only modes with low $\tilde\ell$ and $m$ are likely to be visible using asteroseismology.
As a consequence, we can expect $\sigma$ to be much larger than one.
This means that the dependence of $\omega$ in $m$ can be completely neglected as a first approximation.
Thus, Eq.~\eqref{eq:sep_m} becomes
\begin{equation}
    \Delta\Pi \simeq \frac{2}{m^2}\sqrt{\frac{\left(\tilde\ell+\frac12\right)\pi\int_0^{r_{\rm s}}\frac{N_0}{r}{\rm d}r}{\Omega^3\left(\tilde n+\frac34\right)^3}}.
\end{equation}
This time, for given $\tilde\ell$ and $m$, the spacings scale as $(\tilde n+3/4)^{-3/2}$, and allow one to measure the ratio $\left(\int_0^{r_{\rm s}}N_0{\rm d}r/r\right)/\Omega^3$.

\subsection{Potential targets}

The laws we derived are \emph{a priori} valid when the mode frequencies are much lower than the rotation frequency of the star.
Therefore, such modes and the corresponding period spacings are more likely to be observed in rapidly rotating stars exhibiting \emph{g} modes, such as $\gamma$ Doradus \citep{VanReeth}, SPB \citep{Papics,Moravveji}, or Be stars \citep{Neiner}.
However, to obtain quantitative information about the domain of validity of these expressions, it is necessary to compare them with full numerical computations of modes.

\section{Comparison with the traditional approximation}
\label{sec:comp}

For axisymmetric modes, period spacings similar to those given in Eq.~\eqref{eq:dpa} can be derived within the traditional approximation in the limit of low frequencies \citep[using e.g. Eq.~(29) of][]{Townsend03}.
This suggests that the traditional approximation is valid in this regime.
However, the traditional approximation fails to describe the propagation of waves near the centre of stars, as already mentioned in Sect.~\ref{sec:tradi}.
In our new approximation, in contrast, rays do propagate near the centre because in the eikonal equation~\eqref{eq:eik_gam} the factor $N_0^2$ in front of $k_\theta^2$ in Eq.~\eqref{eq:eik_tra} is replaced by $N_0^2+f^2$ (in front of $k_\gamma^2$).
This difference also appears clearly when looking at the imprint of trajectories in the two approximations on PSS (see Fig.~\ref{fig:lf_lr}).
Both approximations qualitatively agree with the full dynamics far from the centre, but only the low-frequency approximation  (green dashed line) also correctly  describes the behaviour near the centre, where the traditional approximation (blue dash-dotted line) fails.

As seen in Sect.~\ref{sec:ray}, the new invariant is also much better conserved in computations using the full dynamics than the invariant of the traditional approximation tested in Paper~I.
The latter can vary by several orders of magnitude near the boundaries of the resonant cavity, which are not well reproduced by the traditional approximation, and up to a factor two in the bulk of the resonant cavity.

Finally, the traditional approximation is limited to the spherical geometry, whereas our formalism is not, allowing us to account for the centrifugal deformation of stars.

\section{Discussion and conclusions}

\label{sec:dis}

Sub-inertial modes are strongly affected by rotation in the sense that their domain of propagation is limited to a region around the equatorial plane.
Thanks to the present study, we are now able to extract signatures of rotation when its effects are important.
We provide two different period spacings with two different scalings in $I=\int_0^{r_{\rm s}}N_0{\rm d}r/r$ and $\Omega$: one in $(I\Omega)^{-1/2}$ and the other in $(I/\Omega^3)^{1/2}$.
By combining the two, it could be possible to determine stratification (through the integral $I$) and rotation  at the same time, thus constraining mixing processes in rapidly rotating stars \citep[see][]{Moravveji}. 
Besides, for stars whose fundamental parameters are known, the profile of the Brunt-V\"ais\"al\"a frequency can be obtained using stellar models.
Thus, only one of the scalings would be needed to determine the rotation.
For example, this is performed within the traditional approximation for a sample of $\gamma$ Doradus stars by \citet{VanReeth}.

The new predictions for the period spacings found in this paper have been obtained thanks to a certain number of approximations.
In particular, the derivation relies on the assumption that the mode frequencies are much lower than the rotation frequency, and we used the fact that the considered modes are trapped near the equator in
this regime.
However, our model cannot describe some low-frequency modes, such as Rossby and Kelvin modes, because of the form of our eikonal equation (see Paper I).
In addition, we approximated $N_0$, $k_{\rm c}$, and $\zeta$ as functions of the pseudo-radial coordinate $\beta$ alone.
This is probably the limiting factor, since centrifugal deformation induces a stronger dependence of the structure on the latitude when rotation increases towards the critical velocity.
To test the quantitative influence of the approximations we made in this study on the domain of validity of our predictions, we need to confront these predictions to numerical mode calculations, which will be peformed in a forthcoming paper.
Nevertheless, in this study we verified that the approximated dynamics is in qualitative agreement with the asymptotic theory proposed in Paper I by comparing PSS and verifying the conservation of the new invariant.
We found that the full ray dynamics is well approximated up to moderate rotation rates (such as 40\% of the critical velocity).

In the super-inertial regime, our tests suggest that the approximation also gives good results.
However, one has to be cautious because the PSS we showed correspond to the intersection of trajectories with the equatorial plane.
There is no evidence that the approximation correctly describes the dynamics near the poles.
Besides, in this regime, prescriptions for island modes and chaotic modes are also needed.

Because it takes the full Coriolis and centrifugal accelerations into account, the new approximation appears to be more accurate than the traditional approximation, especially near the centre, where it allows for the propagation of waves.

A next step, to go further, would be to introduce differential rotation.
This would modify the domain of propagation of the waves and the associated seismic diagnoses \citep{Mathis09, Mirouh}.
A more general version of the asymptotic theory of Paper I, including the effects of differential rotation, is being built.
If the corresponding ray dynamics also shows integrable structures at low frequency, a generalised version of the low-frequency approximation presented in this paper might be used to derive seismic diagnoses such as period spacings, thus adding new constraints on differential rotation of stars.

\begin{acknowledgements}
V.P. and S.M. acknowledge funding by the European Research Council through ERC grant SPIRE 647383.
The authors acknowledge funding by SpaceInn, PNPS (CNRS/INSU), and CNES CoRoT/\emph{Kepler} and PLATO grants at SAp and IRAP.
The authors thank the anonymous referee for helping to improve the quality of the paper.
\end{acknowledgements}

\bibliographystyle{aa}
\bibliography{refs}

\appendix

\section{Simplification of the canonical eikonal equation}
\label{app:lfa}

The aim of this Appendix is to simplify Eq.~\eqref{eq:eik_pri} into Eq.~\eqref{eq:eik_gam}.
First, Eq.~\eqref{eq:def_pri} can be rewritten
\begin{equation}
    \tan2\alpha=\frac{2N_0\sin\Theta\cos\Theta}{2N_0^2\cos^2\Theta-\left(N_0^2+f^2\right)},
\end{equation}
which, using Eq.~\eqref{eq:ineq}, becomes
\begin{equation}
    \tan2\alpha \simeq-2\frac{N_0^2}{N_0^2+f^2}\sin\Theta\cos\Theta.
\end{equation}
This shows that $\tan2\alpha$ is much smaller than one, and given the fact that $k_\beta=k_s$ at the equator, that $\alpha$ is close to $\pi/2$.
The identity
\begin{equation}
    \tan2\alpha = \frac{2\tan\alpha}{1-\tan^2\alpha}
\end{equation}
can thus be simplified into
\begin{equation}
    \tan2\alpha \simeq -\frac{2}{\tan\alpha}.
\end{equation}
Finally, we obtain
\begin{equation}
    \label{eq:alp}
    \sin\alpha\simeq 1\quad\text{and}\quad\cos\alpha \simeq \frac{N_0^2\sin\Theta\cos\Theta}{N_0^2+f^2}.
\end{equation}

Consequently, we have
\begin{equation}
    \cos(\alpha-\Theta)\simeq\sin\Theta\quad\text{and}\quad\sin(\alpha-\Theta)\simeq\frac{f^2\cos\Theta}{N_0^2+f^2}.
\end{equation}
Equation~\eqref{eq:eik_pri} then yields
\begin{equation}
    \omega^2=\frac{\left[k_\beta^2\frac{f^2N_0^2\cos^2\Theta}{\left(N_0^2+f^2\right)^2}+k_\gamma^2\right]\left(f^2+N_0^2\right)+N_0^2k_\phi^2+f^2\cos^2\Theta k_{\rm c}^2}{k^2+k_{\rm c}^2},
\end{equation}
where we used the fact that $f^2+N_0^2\sin^2\Theta\simeq f^2+N_0^2$.
Besides, in outer layers where $k_{\rm c}$ is dominant, we also expect $N_0$ to be much larger than $f$, and thus $\cos\Theta\simeq\cos\delta$ in the last term of the numerator.
This gives Eq.~\eqref{eq:eik_gam}, as expected.

\section{Relation between $k_\gamma$ and $k_\delta$}
\label{sec:gamdel}

Using Eqs.~\eqref{eq:dgam} and~\eqref{eq:alp}, one can write
\begin{equation}
    \label{eq:dgamcyl}
    {\rm d}\gamma \simeq \frac{N_0^2\sin\Theta\cos\Theta}{N_0^2+f^2}{\rm d}s - {\rm d}z.
\end{equation}
Near the equator, $\theta$ and $\Theta$ are expected to be close to each other.
In spherical coordinates, Eq.~\eqref{eq:dgamcyl} becomes
\begin{equation}
    {\rm d}\gamma \simeq -\frac{f^2}{N_0^2+f^2}\cos\theta{\rm d}r+r\sin\theta{\rm d}\theta.
\end{equation}

We now distinguish between two cases.
First, near the centre, $N_0$ is much smaller than $f$, and thus
\begin{equation}
    {\rm d}\gamma \simeq -{\rm d}(r\cos\theta).
\end{equation}
The definition of $\delta$ in Eq.~\eqref{eq:del} yields
\begin{equation}
    {\rm d}\gamma \simeq -{\rm d}(\zeta\cos\delta),
\end{equation}
where $\zeta$ is defined by Eq.~\eqref{eq:zet} and is nearly constant near the centre, since $N_0$ scales as $r$.
It follows that
\begin{equation}
    \label{eq:dgamddel}
    {\rm d\gamma}\simeq \zeta{\rm d}\delta,
\end{equation}
where we used the fact that $\sin\delta\simeq1$.

Second, when $N_0$ is much larger than $f$,
\begin{equation}
    {\rm d}\gamma\simeq r\sin\theta{\rm d}\theta.
\end{equation}
Using again Eq.~\eqref{eq:del} and the fact that $\sqrt{N_0^2+f^2}/N_0$ is almost constant in the considered regime, one can write
\begin{equation}
    -\sin\theta{\rm d}\theta \simeq -\frac{\sqrt{N_0^2+f^2}}{N_0}{\rm d}\delta,
\end{equation}
which implies Eq.~\eqref{eq:dgamddel}, as before.

In both regimes, we thus have
\begin{equation}
    k_\gamma = \frac{\partial\Phi}{\partial\gamma}=\frac{\partial\Phi}{\partial\delta}\frac{\partial\delta}{\partial\gamma}\simeq\frac{k_\delta}{\zeta},
\end{equation}
as stated in Eq.~\eqref{eq:kdel}.

\section{Derivation of the new invariant}
\label{sec:invar}

We search for an invariant that makes the problem separable in $\beta$ and $\delta$.
This invariant then can be expressed as a function of $\delta$ and $k_\delta$ only $\chi(\delta, k_\delta)$.
The fact that $\chi$ is invariant is written
\begin{equation}
    \frac{\partial\chi}{\partial\delta}\frac{{\rm d}\delta}{{\rm d}t} + \frac{\partial\chi}{\partial k_\delta}\frac{{\rm d}k_\delta}{{\rm d}t}=0.
\end{equation}
This implies that $\chi$ is constant along the characteristics given by
\begin{equation}
    \frac{{\rm d}\delta}{\frac{{\rm d}\delta}{{\rm d}t}} = \frac{{\rm d}k_\delta}{\frac{{\rm d}k_\delta}{{\rm d}t}}.
\end{equation}
Using Eqs.~\eqref{eq:dyndel} and \eqref{eq:dynkdel}, one obtains after simplifications
\begin{equation}
    \frac{\sin\delta\cos\delta{\rm d}\delta}{\sin^2\delta + \frac{N_0^2}{f^2}} = \frac{k_\delta {\rm d}k_\delta}{\zeta^2\left(k_\beta^2+\frac{k_\delta^2}{\zeta^2}+k_{\rm c}^2\right)}.
\end{equation}
Assuming that $N_0$, $\zeta$ and $k_{\rm c}$ do not depend on $\delta$, the integration of this equation yields
\begin{equation}
    \ln\left(\sin^2\delta+\frac{N_0^2}{f^2}\right) = \ln\left[\zeta^2\left(k_\beta^2+\frac{k_\delta^2}{\zeta^2}+k_{\rm c}^2\right)\right] + C,
\end{equation}
where $C$ is constant.

If $C$ is an invariant, any function of $C$ is so.
In particular, we choose $\chi$ as defined in Eq.~\eqref{eq:chi}.
To verify that $\chi$ really is invariant, one must verify that it can also be expressed as a function of $\beta$ and $k_\beta$ only, which requires
\begin{equation}
    \frac{\partial\chi}{\partial\beta}\frac{{\rm d}\beta}{{\rm d}t} + \frac{\partial\chi}{\partial k_\beta}\frac{{\rm d}k_\beta}{{\rm d}t}=0.
\end{equation}
This is easily carried out using Eqs.~\eqref{eq:dynbet} and \eqref{eq:dynkbet}.

\end{document}